\documentclass[sigconf]{acmart}
\usepackage{xcolor}
\AtBeginDocument{%
  \providecommand\BibTeX{{%
    \normalfont B\kern-0.5em{\scshape i\kern-0.25em b}\kern-0.8em\TeX}}}

\copyrightyear{2024}
\acmYear{2024}
\setcopyright{rightsretained}
\acmConference[Websci '24]{ACM Web Science Conference}{May 21--24, 2024}{Stuttgart, Germany}
\acmBooktitle{ACM Web Science Conference (Websci '24), May 21--24, 2024, Stuttgart, Germany}\acmDOI{10.1145/3614419.3643996}
\acmISBN{979-8-4007-0334-8/24/05}

\newcommand{\new}[1]{\textcolor{black}{#1}}
\newcommand{\cameraready}[1]{\textcolor{black}{#1}}

\begin{document}

\title{Echo Chambers in the Age of Algorithms: An Audit of Twitter’s Friend Recommender System}

\author{Kayla Duskin}
\email{kduskin@uw.edu}
\affiliation{%
  \institution{Information School, University of Washington}
  \city{Seattle}
  \state{WA}
  \country{USA}
}

\author{Joseph S. Schafer}
\affiliation{%
  \institution{Dept. of Human Centered Design and Engineering, University of Washington}
  \city{Seattle}
  \state{WA}
  \country{USA}
}

\author{Jevin D. West}
\affiliation{%
  \institution{Information School, University of Washington}
  \city{Seattle}
  \state{WA}
  \country{USA}
}

\author{Emma S. Spiro}
\affiliation{%
  \institution{Information School, University of Washington}
  \city{Seattle}
  \state{WA}
  \country{USA}
}

% \renewcommand{\shortauthors}{Duskin et al.}

\begin{abstract}
The presence of political misinformation and ideological echo chambers on social media platforms is concerning given the important role that these sites play in the public's exposure to news and current events. \new{Algorithmic systems employed on these platforms are presumed to play a role in these phenomena, but little is known about their mechanisms and effects.}
In this work, we conduct an algorithmic audit of Twitter's Who-To-Follow friend recommendation system, the first empirical audit that investigates the impact of this algorithm in-situ. We create automated Twitter accounts that initially follow left and right affiliated U.S. politicians during the 2022 U.S. midterm elections and then grow their information networks using the platform's recommender system. We pair the experiment with an observational study of Twitter users who already follow the same politicians. Broadly, we find that while following the recommendation algorithm leads accounts into dense and reciprocal neighborhoods that structurally resemble echo chambers, the recommender also results in less political homogeneity of a user's network compared to accounts growing their networks through social endorsement. Furthermore, accounts that exclusively followed users recommended by the algorithm had fewer opportunities to encounter content centered on false or misleading election narratives compared to choosing friends based on social endorsement. 
\end{abstract}

\begin{CCSXML}
<ccs2012>
<concept>
<concept_id>10003120.10003130.10003131.10011761</concept_id>
<concept_desc>Human-centered computing~Social media</concept_desc>
<concept_significance>500</concept_significance>
</concept>
<concept>
<concept_id>10003120.10003130.10003131.10003292</concept_id>
<concept_desc>Human-centered computing~Social networks</concept_desc>
<concept_significance>500</concept_significance>
</concept>
<concept>
<concept_id>10002951.10003260.10003261.10003270</concept_id>
<concept_desc>Information systems~Social recommendation</concept_desc>
<concept_significance>300</concept_significance>
</concept>
</ccs2012>
\end{CCSXML}

\ccsdesc[500]{Human-centered computing~Social media}
\ccsdesc[500]{Human-centered computing~Social networks}
\ccsdesc[300]{Information systems~Social recommendation}

\keywords{Social Media, Recommender Systems, Algorithmic Audit}


\settopmatter{printfolios=true}
\maketitle
\section{Introduction}

Social media platforms play an increasingly important role in the consumption and sharing of information and are now a source of news for many \cite{pew_2022}. Growing alongside the dependence on these platforms for news is the recognition that they may facilitate information echo chambers, contribute to political polarization, and exacerbate the spread of misinformation \cite{Cinelli2021-jr, Kubin2021-lc, pew_misinfo}. These issues pose significant risk across a variety of domains, from democratic processes to public health and well-being \cite{Sanchez_Middlemass_2022,Lee_Sun_Jang_Connelly_2022}. Given the prominent role of social media in information infrastructure and the potential negative societal impacts, it is crucial to understand the complex combination of factors -- behavioral and technological -- that may exacerbate the issues found online. 

One key piece of the puzzle is the underlying algorithms that are commonplace on social media sites, from content recommendation and filtering, to serving ads, to friend suggestions. 
There is a growing body of scholarly work attempting to quantify how these algorithms alter the information ecosystem, for example exploring how Twitter's~\footnote{For clarity given recent changes to the social media platform X, formerly known as Twitter, we use the name and terms associated with the site during the time of the study including terms such as tweet, retweet, and follow.} content recommendation algorithm may distort users' experiences in particular ways \cite{Bartley2021-we, Chen2021-tg,Bandy2021-lg,Bandy_Lazovich_2022}. \new{Interest in these algorithms has only increased with Twitter's release of source code for both their content and friend recommendation systems in March 2023~\footnote{https://blog.twitter.com/engineering/en\_us/topics/open-source/2023/twitter-recommendation-algorithm}. While the publication of the code helps to inform the public which factors are considered when ranking content or suggesting friends, this does not translate into understanding how these algorithms embedded in social systems shape (and are shaped by) user experiences over time. Given that key parameters and trained models remain unavailable, and Twitter continues to restrict access to platform data that would allow for outside assessment of the system, several recent articles have pointed out the limited utility of this information \cite{wiredTwittersOpen, atlanticTransperent}.}
Thus, despite progress, there are still vital gaps in our understanding of the impact of these recommender systems across a variety of domains. It has yet to be empirically shown, for example, if (and how) Twitter's friend-recommendation algorithm encourages political echo chambers on its platform. In this work, we seek to address this specific gap by employing an algorithmic audit of Twitter's Who-To-Follow suggestions in order to answer the following questions: 

\begin{itemize}
\item[] \textbf{RQ1:} How, if at all, does the friend recommendation algorithm impact the structural qualities of personal networks of those who use it?
\item[] \textbf{RQ2:} Does the friend recommendation algorithm impact the partisan makeup of resulting personal networks?
\item[]\textbf{RQ3:} Does use of friend recommendations play a role in the amount of false and misleading content to which a user is potentially exposed?
\end{itemize}

We concentrate this study on political polarization and false and misleading election-related content in the United States, running an experiment on the Twitter platform to coincide with the 2022 U.S. Midterm Elections. In this paper, we make the following main contributions: 
\begin{enumerate}
    \item We describe the design and implementation of the first algorithmic audit of Twitter's Who-To-Follow algorithm using automated accounts that mimic a user building a friendship network.
    \item We collect and analyze a unique, longitudinal data set of each account's evolving social network and the content they could have encountered through the social relationships, as well as observational data of comparable real Twitter users. \new{The data includes 1,331,258 tweets from 7,693 unique followed users as part of the audit experiment and 928 users as part of the observational study along with 52,489,602 tweets posted by their friends} 
    \item We present empirical findings that indicate that Twitter's friend recommendation algorithm results in \textit{less} polarized networks than those observed on the platform and when compared to social endorsement based strategies for forming new connections (e.g. following those who existing friends retweet). This result has important implications for understanding the role of social algorithms in both facilitating political polarization and potentially reducing its effects.
\end{enumerate}

%\subsection{Contributions}
    
\section{Background \& Related Work}
This work employs an algorithmic audit of Twitter's friend recommendation system to identify how its features may (or may not) result in a networked informational environment that resembles a political echo chamber. Here, we provide a general overview of prior studies of online echo chambers and polarization; we then introduce algorithmic auditing as a research design, and finally, we focus on recommender systems and the role they may play in polarization and misinformation in online settings. These domains of study inform our work which is, to our knowledge, the first empirical audit investigating the impact of Twitter's friend recommendation algorithm on echo chambers and encounters with false and misleading information within the platform.

\subsection{Echo Chambers and Political Polarization}
\label{sec:echo_chambers}
The metaphor of an echo chamber refers to a group of individuals who share the same opinions or world view and repeat (or \textit{echo}) those homogeneous views within the group, thus reinforcing their shared pre-existing beliefs (or in some cases pushing beliefs to be even more extreme) \cite{Garrett_2009}. A widely held concern is that echo chambers may lead to divergence of public opinion and increase political polarization \cite{Sunstein_2002}. 

Echo chambers are most often identified through both their structural properties (e.g. density and transitivity of social ties, network homophily in regards to identity or opinion) as well as through the \new{ideological homogeneity of their members} (e.g. \new{similarity} of content shared within a group, degree of difference between distinct groups). Many studies on digital trace data have used these methods to characterize ideological echo chambers on a variety of social media platforms, especially in regard to politics \cite{Cinelli2021-jr, Garimella2018}, their relationship to online political polarization \cite{Garimella2017-om, Cota2019-ut, Lee2014-hh,Flaxman_Goel_Rao_2016} and the spread of misinformation \cite{DelVicario2016,Nikolov_Flammini_Menczer_2021}. Observational studies have largely found that users end up largely segregated into communities centered on homogeneous opinions \cite{Terren2021-fj}. That said, there are studies and reports that argue for a tempered interpretation of echo chambers, saying they are less prevalent than often assumed \cite{Guess_2018,Dubois2018-dc}. 

A related but distinct concept is a filter bubble, which is used to describe how a user's exposure to ideologically homogeneous content is influenced by algorithmic systems that filter and suggest content for the user \cite{pariser2011filter}. Echo chambers may exist for a variety of reasons, including psychological phenomena such as confirmation bias or selective exposure; filter bubbles are one potential contributing factor to their existence. However, the link between echo chambers and filter bubbles is not fully understood. In fact, at least on YouTube, it appears that algorithmically suggested content is not the primary driver of the consumption of extreme content \cite{Hosseinmardi2021-nc, Chen2022-bh, Ribeiro2023-kc}.

\begin{figure*}[t]
\centering
\includegraphics[width=1.8\columnwidth]{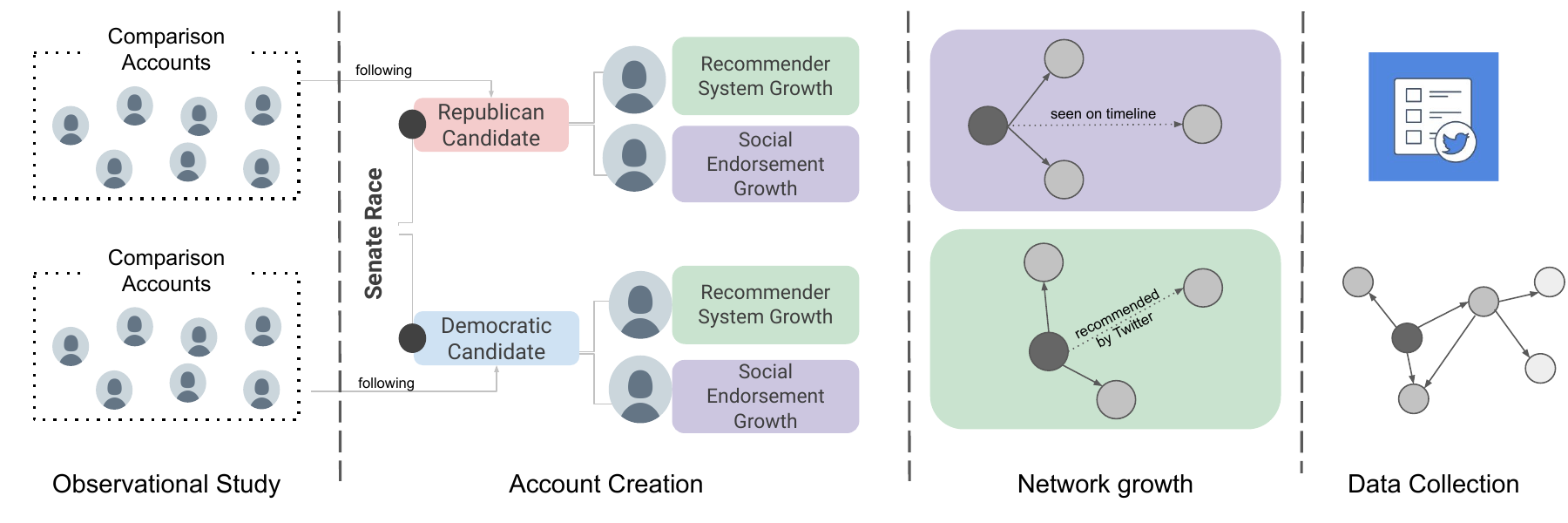} 
\caption{Audit accounts were created in matched pairs to compare network growth strategies where both accounts initially followed a single U.S. politician (seed account) running for office during the 2022 midterms. Seed accounts were chosen as the Democratic and Republican candidates from five Senate races. Across the five Senate races, 20 accounts were created in the audit (two candidates per race, two audit accounts per candidate).}
\label{fig:account_schema}
\end{figure*} 

\subsection{Algorithmic Auditing}

Algorithmic audits seek to uncover the potential negative impacts of computer algorithms by observing input and outputs to the system \cite{sandvig2014auditing}, and are particularly useful when the details of the system implementation are hidden from the researcher -- as they are for most social media recommender systems. To gain insight into an algorithm's impact on users, scholars have used automated agents, or sock-puppet audits, to simulate a human's experience interacting with the system under study \cite{Diakopoulos2021-dq}.  

Several audits using automated accounts have taken place on Twitter, including a study that found that the initial selection of friends impacts the density, transitivity, and political homogeneity of an account's personal network but found no consistent evidence of political bias in the chronological news feed \cite{Chen2021-tg}. Bandy and Diakopoulos \cite{Bandy2021-lg} found that Twitter's algorithmically curated "Home Timeline" increases links to external sites and increases source diversity while potentially increasing the disparity in partisan differences when compared to the reverse chronological feed. In contrast, Bartley et al. \cite{Bartley2021-we} found that Twitter's algorithmic timeline increases inequality in exposure to friends' tweets. 

Motivated by concerns of ``rabbit holes" that lead to increasingly fringe content, YouTube's recommendation algorithm has also been the topic of debate and study. Through a study using automated accounts, Brown et al. \cite{Brown2022-ai} found that YouTube's recommendation algorithm pushes users into mild political echo chambers, and biases toward moderately conservative content. Several studies \cite{Hussein2020-tu, Papadamou2022-ym} confirm YouTube's filter bubble effect, where watching videos that promote misinformation triggers recommendations for similarly misinformative content. Similarly, it has been shown that YouTube's recommendations direct right-leaning users into increasingly radical content \cite{Haroon2022-ks}.

\subsection{Recommendation Systems and Echo Chambers}
Friend recommendation systems, where an algorithmic system suggests accounts that a user may be interested in connecting with, are now ubiquitous across social media platforms. When it was introduced in 2010, Twitter's "Who to Follow" friend recommendation algorithm increased follower counts across all users, but resulted in a larger boost for already popular accounts \cite{Su2016-ry}. Work to understand how this system may impact polarization or exposure to false and misleading information has so far been limited to simulations. One such simulation found that recommenders that connect structurally similar nodes (those that already share neighbors) result in network topology that enables opinion polarization \cite{Santos2021-vm}. \new{Through an analysis of five recommendation algorithms on a synthetic network,  Ferrara et al. \cite{Ferrara2022-ui} found that these systems have the potential to reduce the visibility of minority groups in bi-populated networks.} Cinus et al. \cite{cinus2022effect} found that friend-recommenders can significantly contribute to echo chambers, but only if the initial network is sufficiently homophilous. Using a data set of COVID-19 related misinformation, Tommasel and Menczer \cite{Tommasel2022-uv} found that friend recommendation algorithms that prioritize diversity of connections recommend more misinformation-spreaders and that effect of the recommendation algorithm is greater when misinformation-spreaders are already well connected within the network.  \new{The simulation-based research summarized above provides valuable insight into how friend recommenders have the \textit{potential} for impact on misinformation and ideological polarization, however the function and impact of these systems -- as they exist in the wild -- is understudied.}

\section{Data and Methods}

The audit described in this study, and concurrent data collection, was designed to probe the impact of Twitter's friend recommendations on both the characteristics of the followed accounts and the content those accounts produce through the use of automated accounts. The resulting data set, as such, encompasses timestamped observations of networks (i.e. the accounts followed), the characteristics of these social contacts, as well as the content produced. We pair the data collected on the automated accounts with an observational study of comparable accounts already existing on Twitter. In this section, we review the platform under investigation, present the audit design and implementation, and discuss the resulting data and the methods used to estimate key features from the data, including identifying false and misleading content in posts and estimating account partisanship. This study was reviewed the by the University of Washington Human Subjects Division and determined not to involve human subjects. Therefore, review and approval by the IRB was not required.

\subsection{Platform Overview}

This work was conducted on Twitter from September through December 2022. The choice of platform was motivated by three primary factors: (1) its nearly ubiquitous use in political discourse online, (2) its reliance on algorithmically determined recommendations of social relationships, and (3) its ease of access in terms of data collection (at the time of this study). Twitter's model of offering both suggestions for new connections is common practice across many popular social media platforms and, therefore, can be viewed as illustrative of systems of this kind. We recognize that social media platform affordances and algorithmic choices are not fixed, but rather ever-evolving. Thus, while the data collection period for this study preceded a known phase of volatility of Twitter's algorithmic recommendation system \cite{Rohlinger2023-fg}, this work is limited to the platform as it existed during the observation period. In offering a detailed snapshot of one social media platform, our hope is that the results can provide insight into algorithmic effects on other platforms. 

\subsection{Audit Design and Implementation}
\label{sec:audit_design}
To reveal qualities of an otherwise opaque algorithmic system, it is useful to mimic how a human might engage with that system \cite{Diakopoulos2021-dq}. \new{We choose to employ this approach for our study because, while simulation studies conducted on synthetic or samples of data have probed similar research questions, an algorithmic audit can reveal the outcomes of the real system as a whole as it operates on the massive underlying network of the platform.} 
In this study, we seek to isolate the influence of algorithmic recommendations by creating automated accounts (hereafter referred to as \emph{audit accounts}) that navigate Twitter according to predetermined protocols. The audit consisted of three stages: account creation, network growth, and content collection (depicted in Figure~\ref{fig:account_schema}). To accomplish this work, we used Selenium\footnote{\url{https://selenium-python.readthedocs.io/}} and Python to schedule daily actions for each account and collected account metadata and content through Twitter's academic API.

\subsubsection{Audit Account Creation.}

We created twenty Twitter accounts and initiated each as if it were a new user of the platform interested initially in a single candidate running for U.S. Senate, which we refer to as the \emph{seed candidate}; in other words each new account enters the system and forms a connection to a single existing account of interest. We chose five U.S. Senate races out of the ten races predicted to be the closest on the 2022 Senate Election Forecast published by the popular and widely recognized FiveThirtyEight blog\footnote{https://projects.fivethirtyeight.com/2022-election-forecast/senate/}. For each Senate race, we created four unique accounts -- two which initially followed the Republican candidate and two initially following the Democratic candidate. 

The audit accounts were then assigned to one of two conditions to grow connections over time. The first pair (i.e. Democratic/Republican seeds) of accounts followed a network growth protocol based on the friend recommender system, where network ties are added based on Twitter suggestions. We refer to these audit accounts as recommender-system (RS) audit accounts. \new{The purpose of the RS audit accounts is to demonstrate networks created solely by relying on the friend recommendation algorithm to add new connections. To put those results in context, we also create a set of accounts that ignore the friend recommendation algorithm and instead grow their networks based on the implicit recommendations of existing friends, similarly to how Chen et al. \cite{Chen2021-tg} grow the networks of their neutral drifter bots.} This other network growth protocol selects new accounts to follow based on social-endorsement, where connections were added to accounts retweeted by existing connections. We refer to these accounts as social-endorsement (SE) audit accounts. The two network growth protocols are discussed in more detail in the following section. A schema of the account creation, across races and network growth conditions, is shown in Figure~\ref{fig:account_schema}.

To avoid unintentional influence on Twitter recommendation algorithms, as well as the actions of other accounts in the Twitter ecosystem, the audit accounts were given nondescript names and opted out of selecting any ``interests" when prompted at the time of creation. Additionally, profile descriptions, photos, and banners were left blank to avoid any amount of deception or influence on how others may perceive the account.  

\subsubsection{Network Growth.}
\label{sec:net_growth}

Once created, each audit account entered a network growth phase, following a set of specific steps depending on the assigned network growth protocol. This phase began on September 21, 2022 and continued for each account until it reached a predetermined threshold for the number of friends. This threshold was chosen based on matching the median number of friends in the networks of the corresponding comparison groups (See Section~\ref{sec:comparison_groups}). Audit accounts were in the network growth phase for three to eight weeks, depending on their stopping conditions. 

\noindent \emph{Recommender-system (RS) network growth:} RS accounts logged on twice per day, approximately 12 hours apart starting at 6:00 am PST. The RS audit account then followed all users that were suggested in the "Who to Follow" window on Twitter's home page. Typically this consisted of two to three accounts and this process was done twice during each session by following all accounts suggested on the home page and then refreshing the page and repeating the process, meaning that four to six new accounts were followed twice per day. Promoted accounts were ignored during this process. 
 
\noindent \emph{Social-endorsement (SE) network growth:} SE accounts also conducted a series of following actions twice per day, approximately 12 hours apart starting at 6:00 am PST. In this condition, the 200 most recent tweets from the audit account's friends were collected using the Twitter API's home timeline call. We considered all users that had been retweeted by an audit account's friends as potential new friends, and randomly selected six accounts to follow, weighted by how often that account appeared (i.e., was retweeted) in the most recent 200 tweets. 

\subsubsection{Account Behavior}
\new{Throughout the duration of the study, the audit accounts did not tweet, like, quote, retweet content or otherwise engage with existing accounts in any way other than following them. We deploy the audit accounts as passive consumers, rather than as creators, of content for several key reasons. First, it has been shown that activity on the platform is highly skewed, with the top (i.e. most active) 25\% of users producing 97\% of posts \cite{mcclain2021behaviors}. Importantly, the bottom 75\% of users produce a median of 0 posts (including original tweets, retweets, and replies) per month \cite{mcclain2021behaviors}, making inactivity the norm that we aim to replicate. Second, we consider the risk of amplifying harmful content or deceiving other users on the platform to be too great to justify content engagement. Given the context of the 2022 US midterm elections, programming the audit accounts to retweet or like popular or recent tweets from their timeline (as in \cite{Chen2021-tg}), could have inadvertently led to sharing or boosting the prominence of misleading election information during a critical time period, an outcome antithetical to our work. Withholding from engagement actions fixes this input to the recommendation algorithm across all of the audit accounts while other aspects (e.g. the local network of each audit account) vary.}  

\begin{table*}
\begin{center}
    \caption{Mean values for each statistic described in Section \ref{sec:results_rq1}. The first five measurements are aggregated to form a representative metric for each audit account's personal network, then we take the mean across each group. Remaining measurements produce a single value per account network and the mean across each group is reported here. Unreported measurements in the are due to the lack of complete data to measure the network structural properties of the comparison accounts.}
\begin{tabular}{lrrr|rrr}
\toprule
Seed Candidate Party & \multicolumn{3}{c}{\textbf{Democrat}} & \multicolumn{3}{c}{\textbf{Republican}} \\
Group & Recommender &   Social &   Comparison  & Recommender &   Social &   Comparison \\
\midrule
median friends           &     9222.00 &   1867.40 &    1668.34 &     6782.40 &   2742.20 &    1153.18 \\
median followers         &   396325.88 &  21868.60 &  420799.92 &   232321.10 &  29193.70 &  252467.15 \\
median account age (years)       &        6.76 &      9.39 &      10.88 &        8.21 &      7.99 &      10.65 \\
median tweets per year       &     2400.94 &   4416.42 &    1478.38 &     1919.57 &   5228.45 &    1430.82 \\
mean percent verified      &       31.38 &     30.43 &      52.13 &       36.67 &     30.55 &      50.93 \\
\midrule
E-I Index           &       0.31 &    -0.02 &        0.07 &       0.61 &     -0.58 &       0.15 \\
density               &        0.18 &      0.07 &        - &        0.19 &      0.13 &        - \\
reciprocity           &        0.77 &      0.61 &        - &        0.72 &      0.61 &        - \\
connected components (weak)   &       14.00 &     25.80 &        - &       17.80 &     31.20 &        - \\
connected components (strong) &       43.25 &     87.60 &        - &       47.20 &     75.20 &        - \\
\bottomrule
\end{tabular}

\label{table:networks}
\end{center}
\end{table*}

\subsubsection{Data collection.}
\emph{Network Observation:} We recorded the following actions for each audit account, noting which accounts were followed and when. This creates a time-stamped record of the network growth of each audit account. Additionally, to get the full picture of each audit account's social neighborhood, we used the Twitter API to collect metadata about each followed account (e.g. the age of the account, whether they were considered to be a verified user, how many friend and followers they have) as well as each followed account's outgoing social ties (e.g. their friends). 

\noindent \emph{Content Collection:} Overlapping with the period of network growth, we collected content for each audit account. To collect all tweets that an account \textit{could} have been exposed to, we used the Twitter API to archive all tweets posted by all of the audit account's friends at regular intervals. This archival process ran approximately every three days for the duration of the content collection period and resulted in 1,322,414 unique tweets during the period of interest.  

\subsection{Observational Study}
\label{sec:comparison_groups}
To establish a baseline for what the personal network of a typical follower of each seed candidate might look like, we defined a comparison group of existing Twitter users for each seed candidate. For each of the political candidate seed accounts, a group of 100 users were randomly sampled from their most recent 2000 followers. After sampling, some accounts were deleted, suspended, or made private resulting in a group of 928 users split into 10 groups (five Senate races with two candidates each) by seed account that serve as a comparison sample of real followers of the political seed candidates. We calculated the median number of friends of each comparison group and used this as the stopping condition for building the audit accounts' networks. \cameraready{That is, we stopped growing the audit account's network when it reached the median network size of the corresponding comparison group.} For each of these followers, we used the Twitter API to collect a snapshot of their personal network (i.e. all of the accounts they are following). Once we identified all of the friends of each comparison account, we retrieved account metadata for each friend account. We also collected the content that comparison accounts were potentially exposed to by retrieving all tweets posted by the comparison accounts' friends. We limit this to a six day period surrounding the midterm election date (November 6th through November 12th) due to the large amount of data this produced. This resulted in a set of 52,489,602 unique tweets.     

\subsection{False and Misleading Content}
\label{sec:misinfo_data}
We enrich the data described above to understand the prevalence of tweets pertaining to false and misleading election narratives. Three of the authors were involved in a real-time, multi-institution collaboration to monitor and analyze rumors and misinformation about the 2022 U.S. midterm election \cameraready{through the Election Integrity Partnership\footnote{\url{https://www.eipartnership.net/}}}. This process involved active monitoring of election-related content across multiple platform to identify emerging unproven claims as they gained popularity online. Once identified, a team of researchers used quantitative assessment of Twitter data (provided by keyword queries using Twitter's V1 streaming API) and qualitative assessment of claims and relevant media (news articles, fact-checks, press releases, etc.) to determine the legitimacy of claims and define the key narratives present. The outcome of this work is a curated set of election-related incidents that contain false or misleading information as well as corresponding keyword-based queries crafted to identify tweets related to each incident in a process similar to \cite{kennedy2022repeat}. \cameraready{The data has been made publicly available \cite{Schafer2024-dl} and a manuscript further detailing this dataset is in progress at the time of publication.} The scope of these incidents and data is narrow -- \cameraready{consisting of false, misleading, or unsubstantiated claims} related to attempts to suppress voting, reduce election participation, confuse voters, or delegitimize election results without evidence. We \cameraready{cross-reference the tweets posted by the friends of the audit accounts and comparison accounts with the dataset of election rumors}, thus identifying a subset of the tweets that each account had the potential to see that pertain to false or misleading election narratives; during the time period of interest (November 6th through November 12th), we identify \cameraready{187,689} such tweets. 

\subsection{Account Partisanship}
\label{data:partisanship}
To \new{approximate} the political partisanship of each audit account's personal network, we make use of an exogenous data set to identify left and right wing political influencers and then label friend accounts within our data set based on how frequently they retweet the identified influencers. 
To do this we make use of a collection of approximately 217 million tweets mentioning English-language terms related to the midterm election, such as `election' and `vote,'; these data capture a broad sample of election-related Twitter conversations and were made available though the election monitoring project introduced in section \ref{sec:misinfo_data}. These posts were then processed using a coengagement network approach \cite{Beers_Schafer_Kennedy_Wack_Spiro_Starbird_2023}, to find communities of accounts that could be considered political influencers based on the size of their engaged audience. 
Through community detection and qualitative analysis of the content they share, two distinct sets of influencers -- one left-leaning, one right-leaning -- are identified for a combined total of 6,721 users. 

We then use a form of label propagation to assign \new{labels} to the users within the audit data set (e.g. the accounts followed by the audit accounts or followed by the comparison accounts). For each account within our data set, we consider the proportion of their election-related retweets that are left or right leaning influencers. If 80\% or more of their retweets are from one of the two parties, the account receives the corresponding partisan label. For example, if an account in our data set retweeted right-leaning influencers nine times and left-leaning influencers one time, it would have a partisan score of 90\% right-leaning and be assigned a right-leaning label. If an account either does not retweet political content \textit{or} retweets left and right wing influencers relatively equally (retweets each party less than 80\% of the time), the account receives a neutral label. The 80\% threshold was validated through qualitative analysis \new{where two of the authors qualitatively coded samples of 50 random tweets from each group and verified that for the left and right-leaning groups, all political tweets matched the assigned party and the majority of tweets were political tweets while for the neutral group the majority of tweets were apolitical or non-partisan}. We found that this method allowed us to label a comparable percentage of users to other common partisanship-labeling methods such as hashtag-based or link-based classification with the benefit that even users who do not share links or hashtags can be labeled based on their retweets.

\subsection{Measuring Echo Chambers}
As discussed in \ref{sec:echo_chambers}, echo chambers are identified using both network measurements and analysis of ideological homogeneity. We construct personal social networks where each node is an account and each directed edge indicates a following relationship for each audit account and comparison account. We measure structural network properties, such as density and reciprocity and the number of strong and weak connected components in the network. Since the edges formed by our audit accounts are artifacts of the study, we remove them before computing these statistics. In terms of analyzing ideological uniformity, we utilize the account-level partisanship labels and tweet-level misinformation labels to quantify the political homogeneity and quality of information experienced by each account.

\subsection{Challenges}
As expected, we encountered several challenges in deploying automated accounts. First, we were limited in the number of accounts we could create due to the difficulty in authenticating many unique accounts. We therefore limited the study to twenty automated accounts (this is above or on par with similar studies, see e.g. \cite{Chen2021-tg,Bandy2021-lg,Bartley2021-we}). Second, during the phase of network growth, some of the accounts were subject to temporary or permanent freezes on adding new friends due to inadvertently triggering Twitter's moderation policies. In these cases, we resumed network growth as soon as possible, but as a result there is non-uniform network growth across the automated accounts. Finally, one of the twenty accounts was suspended during the process and were unable to be reinstated, resulting in incomplete data for those accounts.

\section{Results}
Overall, we find that following Twitter's friend recommendation algorithm results in networks that differ from both comparison accounts and from networks created through social endorsement-based growth. \new{Where possible, we compare all three groups to put the RS audit accounts in full context by comparing to both ``in-the-wild" Twitter users and to the SE audit accounts that match the RS accounts in terms of their age and general behavior but follow accounts endorsed by their friends.} Specifically, we find that utilizing the friend recommender system results in accounts whose personal networks \textit{structurally} resemble echo chambers (RQ1) but are less politically homogeneous (RQ2) and share less content linked to false and misleading election narratives (RQ3), compared to other groups.  

\begin{figure}[t]
\centering
\includegraphics[width=1\columnwidth]{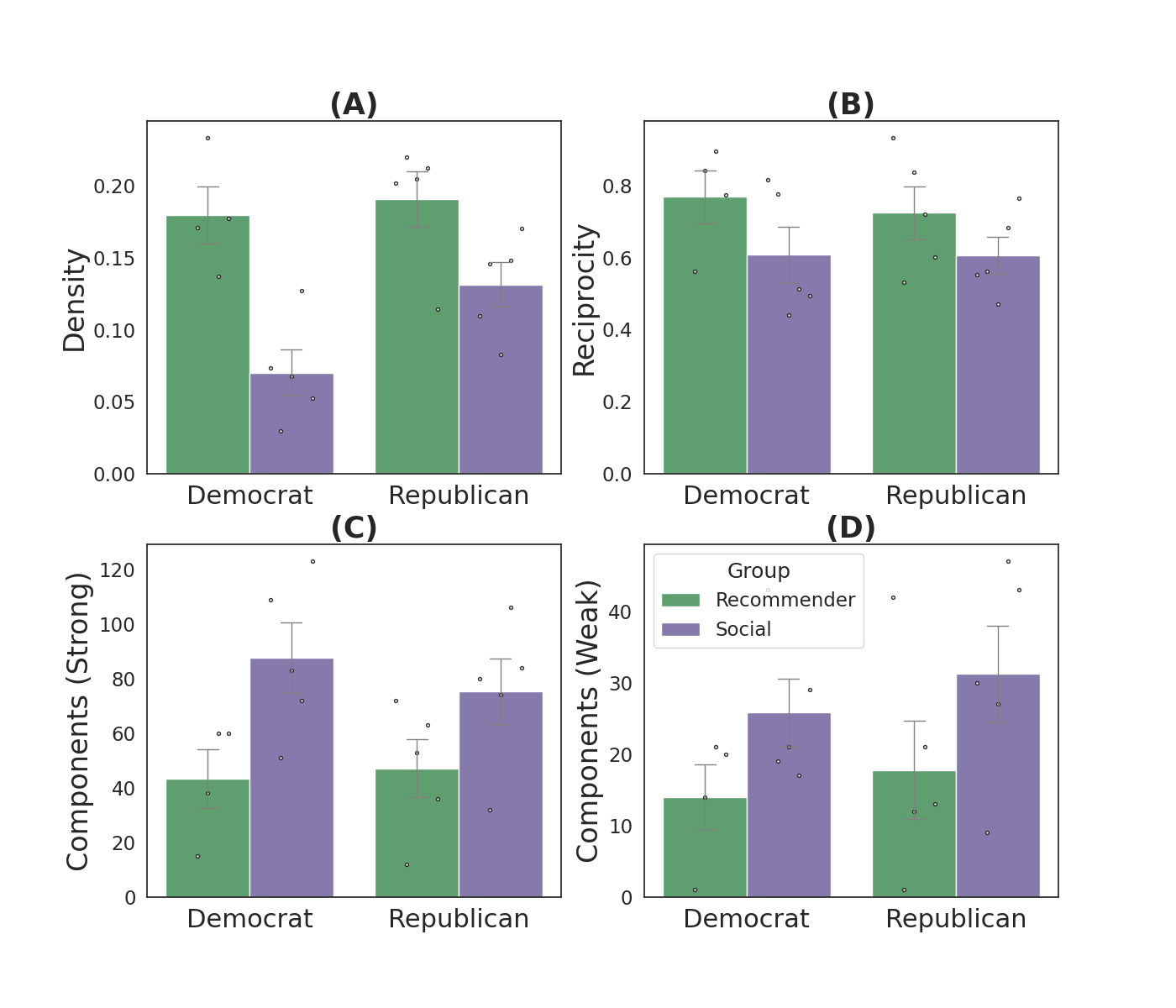} 
\caption{Network measurements of the audit accounts' personal networks, grouped by network growth method and seed party. A) Density B) Reciprocity C) Number of strongly connected components after removing the ego node D) Number of weakly connected components after removing the ego node. Error bars represent the standard error (n=5 audit account networks in each group except the Democratic Algorithmic group where n=4).}
\label{fig:nw_stats}
\end{figure}

\begin{figure}[t]
\centering
\includegraphics[width=1\columnwidth]{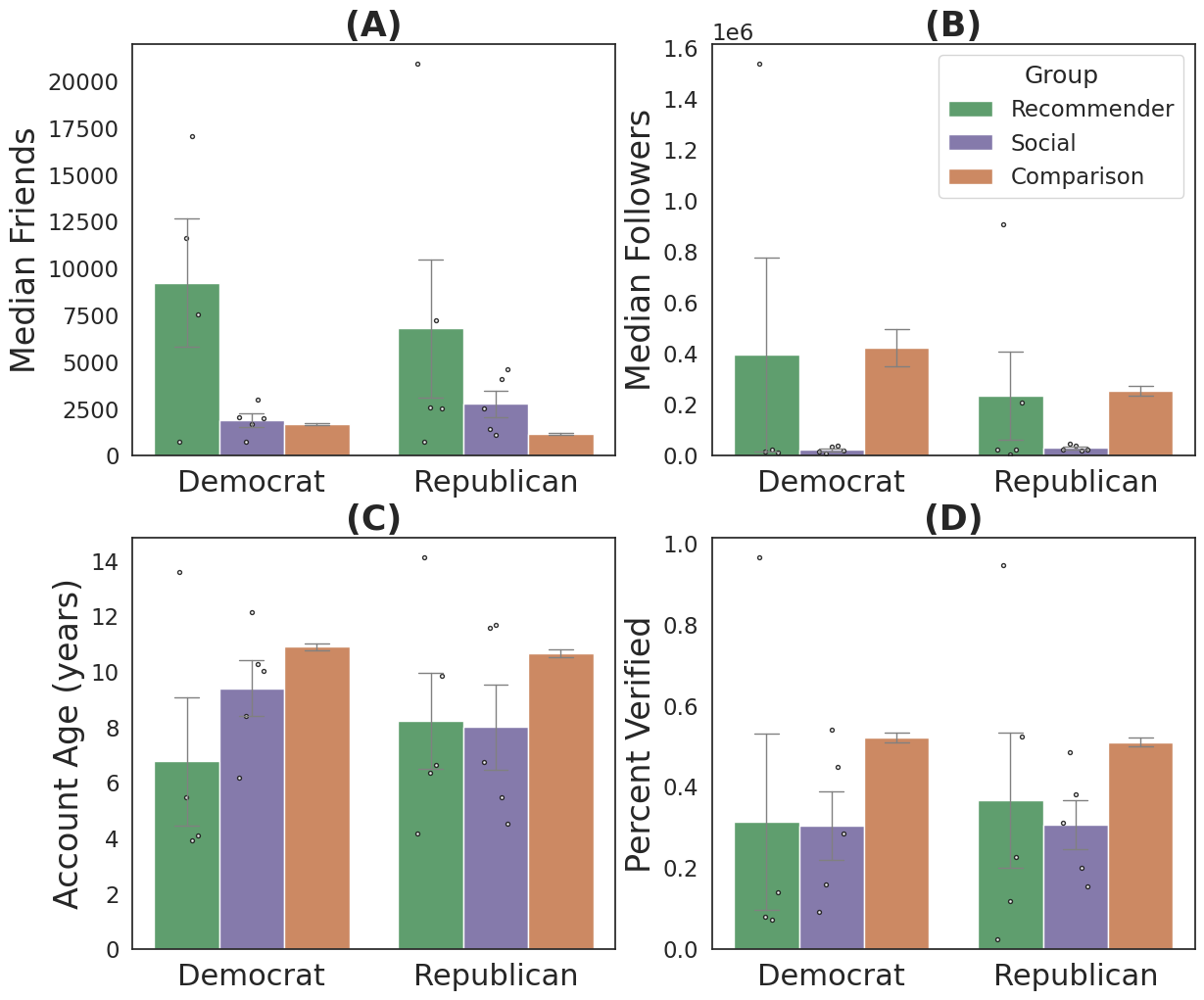} 
\caption{Characteristics of the accounts followed by the audit accounts, grouped by network growth method and seed party. A) Median number of friends B) Median number of followers C) Median account age in years D) Percent of accounts that are verified by Twitter. Error bars represent show the standard error (n=5 audit account networks in each group except the Democratic Algorithmic group where n=4).}
\label{fig:nw_chars}
\end{figure}

\subsection{Comparing Personal Networks (RQ1)}
\label{sec:results_rq1}

We first describe the structural features and characteristics of neighbors present in the resulting personal networks of the audit accounts, comparing across experimental conditions and to the comparison group. These results are illustrated in Figure \ref{fig:nw_stats} and the descriptive statistics for each group are in Table \ref{table:networks}. Structurally we find that recommender-system (RS) accounts end up in highly connected neighborhoods when compared to social-endorsement (SE) accounts. The personal networks of the RS accounts are more dense and more reciprocal than the SE networks. The networks are also less fragmented, as measured by the number of connected components present in the network after removing the ego node (the audit account). RS accounts' personal networks had fewer strongly connected components and weakly connected components compared to the networks of the SE accounts.  We see these patterns across both political parties. 

We also look at the characteristics of the accounts followed by each of the audit accounts and compare them to accounts followed by comparison accounts, as shown in Figure \ref{fig:nw_chars}. We find that the RS accounts follow users that are more popular (i.e. have more followers) when compared to SE accounts but are about equal with comparison group. The RS accounts also follow users who have more friends (out-going ties) compared to the SE accounts and comparison groups. In contrast, the Twitter users from the comparison groups follow a higher percentage of verified accounts than either the RS accounts or SE accounts. \new{Comparison accounts followed older accounts on average than either RS or SE accounts.}

\subsection{Political Homogeneity (RQ2)}
\begin{figure}[t]
\centering
\includegraphics[width=1\columnwidth]{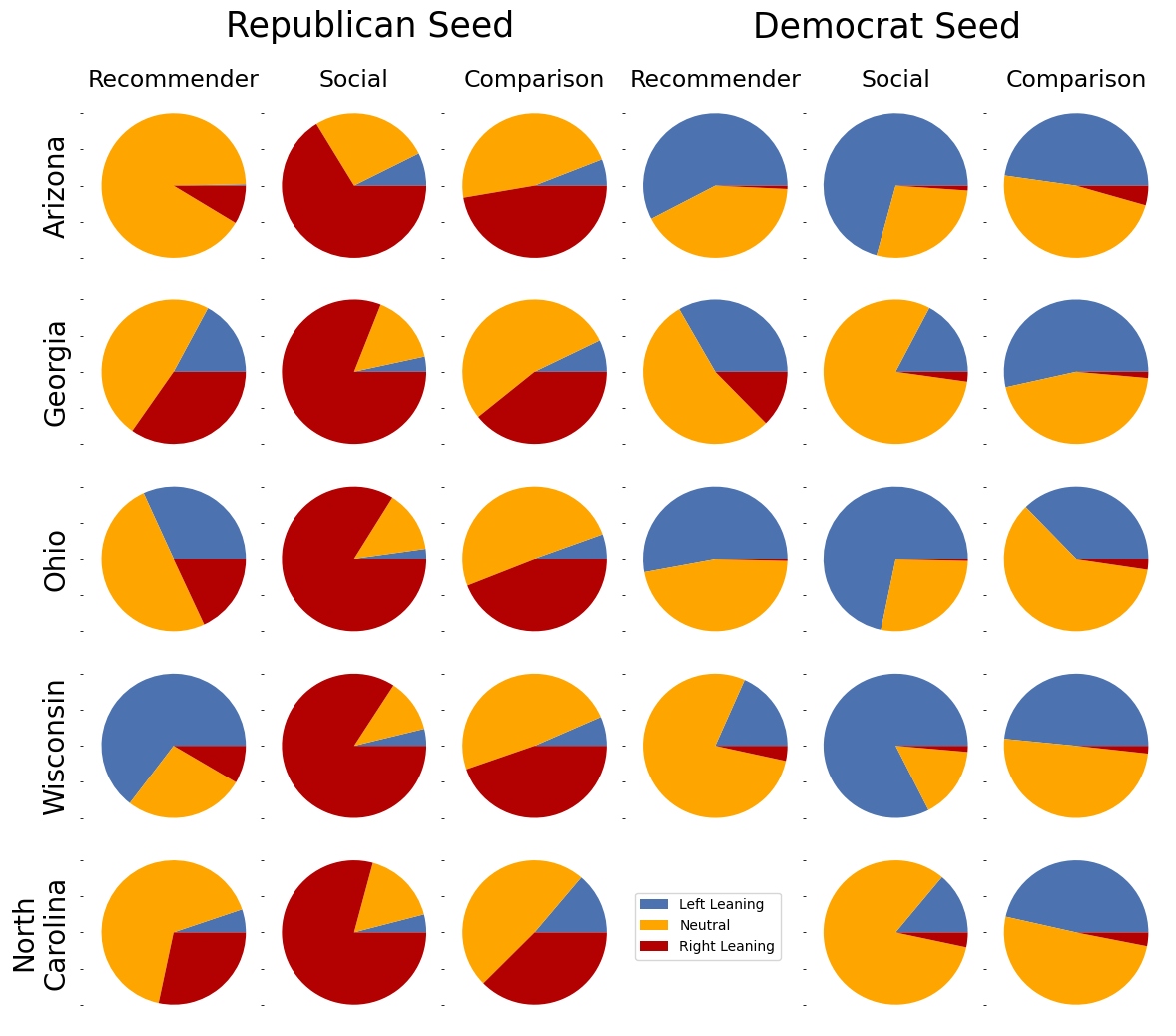} 
\caption{Political makeup of the of the personal network of each audit account (columns 1, 2, 4, and 5) and each comparison group (columns 3 and 6). Users are classified as left-leaning (blue), right-leaning (red) or neutral (orange) as described in section \ref{data:partisanship}}
\label{fig:partisan_pie}
\end{figure}

We analyze the partisanship make up of each audit account's neighborhood using the labels described in Section~\ref{data:partisanship}. We find that the networks of SE accounts display the highest level of political homogeneity, aligning with the party of the seed account -- and that this effect was stronger on the right than on the left as shown in Figure~\ref{fig:partisan_pie}. All of the SE accounts seeded with Republican candidates resulted in personal networks that were dominated by right-leaning accounts, with a mean of 78.91\% of their neighbors identified as right-leaning. Democrat-seeded SE accounts ended up with a mean of 51.23\% left-leaning friends. In contrast, the friend recommendation algorithm resulted in the most politically diverse networks, especially for accounts seeded with a Republican candidate. Republican-seeded RS accounts resulted in an average of only 19.63\% right-leaning friends, while for left-seeded RS accounts 40.53\% of their final network neighbors were left-leaning. The baseline comparison groups showed less variance across candidates in terms of their political makeup and fell between the RS and SE networks in terms of homogeneity, with the baseline comparison groups for Republican candidates averaging 42.52\% right-leaning friends while for the comparison groups for Democratic candidates averaged 46.69\% left-leaning friends. \new{We calculate the Krackhardt EI-Index~\cite{krackhardt1988informal} as a measurement of the comparative frequency of external (inter-party) and internal (intra-party) ties for each ego network and find that the RS accounts have networks exhibiting higher EI-Index values, indicating greater proportions of cross-party ties (see Table \ref{table:networks})}

\begin{figure}[t]
\centering
\includegraphics[width=\columnwidth]{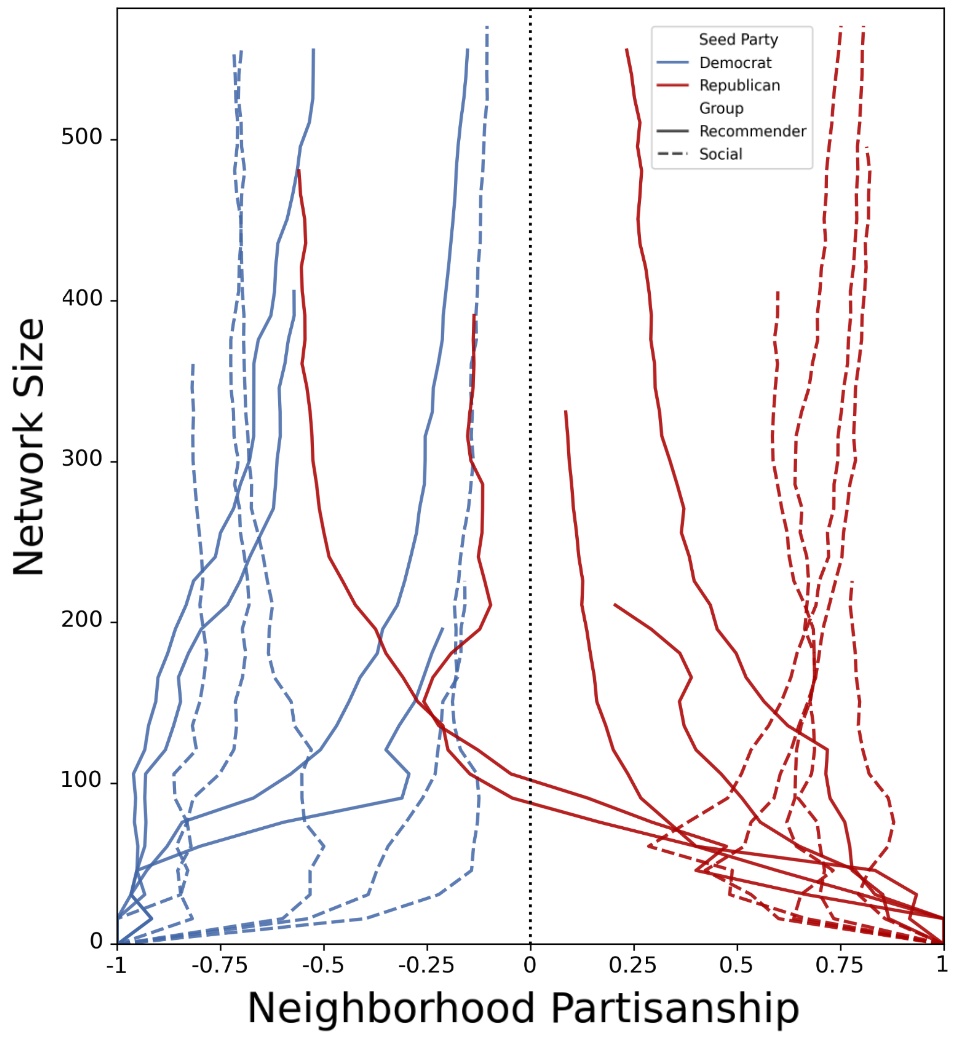} 
\caption{The average partisanship over time of each audit account's set of friends. Left leaning accounts are assigned the value of -1, neutral or apolitical accounts have a value of 0 and right leaning accounts have a value of 1.}\label{fig:partisan_temporal}
\end{figure}

We conducted a temporal analysis of partisanship to asses how each audit account's network neighborhood evolved over time. In most cases, the algorithmic recommendations pulled both left and right seeded accounts away from their baseline party and toward a more ideologically balanced personal network over time (shown in Figure \ref{fig:partisan_temporal}). This effect appears to be stronger for accounts initiated by following a Republican candidate, with two of the audit accounts eventually following more left-leaning than right-leaning accounts. Networks grown through social endorsement tell a different story. While all SE networks moved toward neutrality initially, the right-seeded SE accounts uniformly ended with a strong majority of right-leaning friends. 

\begin{figure}[t]
\centering
\includegraphics[width=\columnwidth]{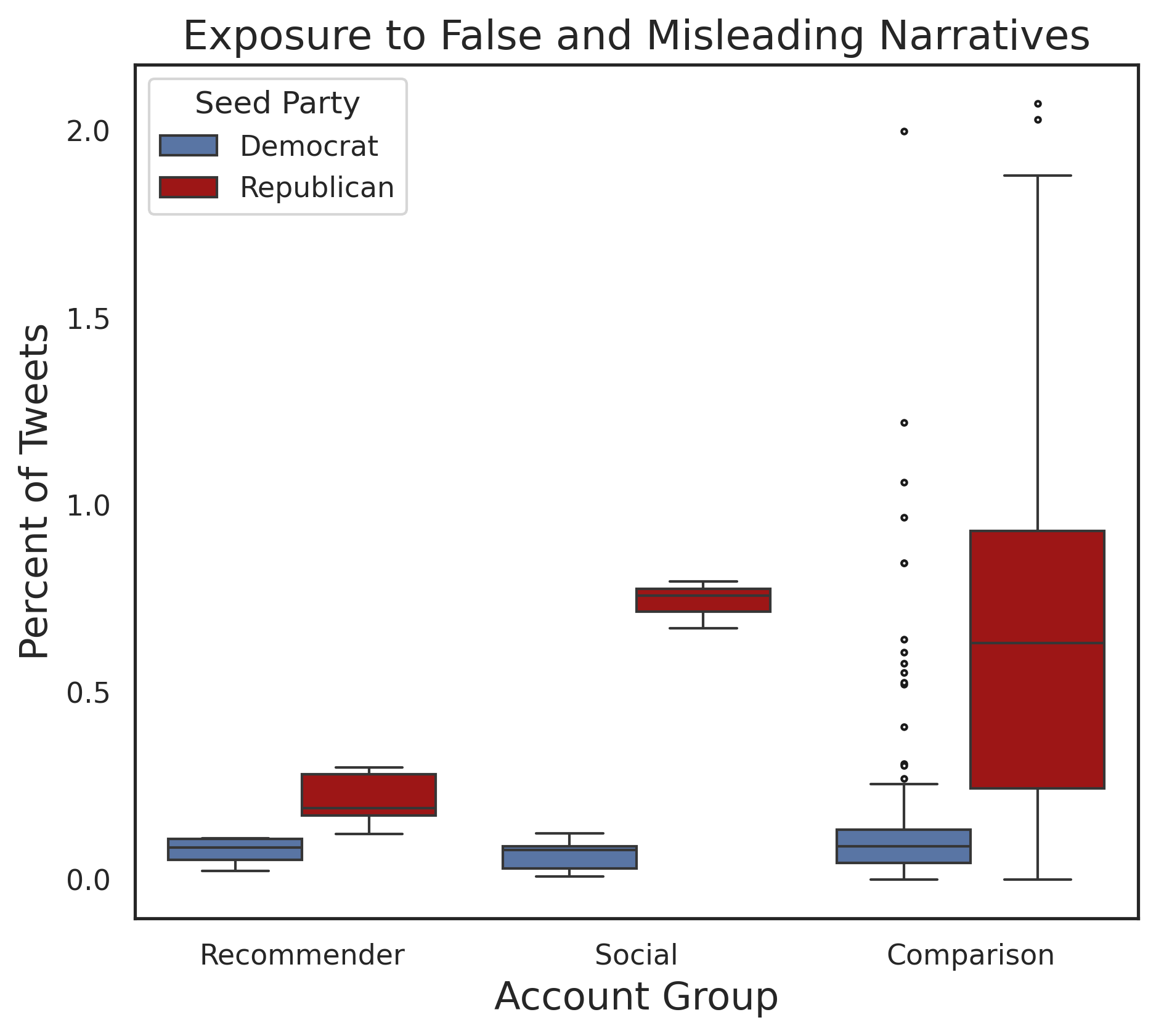}
\caption{Prevalence of tweets about false or misleading election narratives as a percentage of total tweets an account had the potential to be exposed to (all tweets posted by their friends) during the week of the election. }
\label{fig:misinfo}
\end{figure}

\subsection{Potential Exposure to False and Misleading Content (RQ3)}

Figure \ref{fig:misinfo} shows striking partisan differences in the volume of tweets about false or misleading narratives potentially seen by each group. We compute the number of tweets posted by each account's friends during the week of the midterm election that are identified as pertaining to a false or misleading election-related narrative (see \ref{sec:misinfo_data}). We normalize by the total number of tweets each account had the potential to see (the total tweets posted by the accounts they were following) since the number of friends and friend activity level varies. Overall, tweets relating to misleading election content made up only a small portion of potentially viewed tweets. The RS accounts had the lowest potential exposure to misleading narratives. On average, each of the RS accounts was potentially exposed to 44.5 tweets, or 0.076\% of tweets for Democrat-seeded accounts and 102.4 tweets or 0.21\% for Republican-seeded accounts. The SE Democrat-seeded accounts were exposed to 66.6 tweets or 0.065\%, which is similar to the Democrat-seeded RS accounts. However, Republican-seeded SE accounts had the potential to see significantly more misleading content than their RS counterparts -- 1439.4 tweets or 0.74\%. The comparison groups for Democratic candidates were exposed to 342.0 tweets on average or 0.11\% while the comparison groups for Republican candidates were exposed to 1174 tweets on average, or 0.62\%.

In all three groups (RS, SE, and comparison) Democrat-seeded accounts were exposed to fewer tweets about false or misleading narratives. For Democrat-seeded accounts, the form of network growth makes little difference while for Republican-seeded accounts there are clear differences between the RS, SE and comparison groups. For Republican-seeded accounts, the potential exposure of all of the RS accounts fell below the median of the comparison group while all of the SE accounts fell above the median.

\section{Discussion}

\subsection*{Measuring Echo Chambers}
The results of this audit study confirm previous work that has found that early choices in following relationships impact many aspects of a user's experience on a platform \cite{Chen2021-tg,Bandy2021-lg}. We find that Twitter's friend recommendation algorithm leads the automated accounts to neighborhoods that are dense and reciprocal, structurally fitting the criteria often used to define echo chambers. However, analysis of ideological partisanship shows that the friend recommendation algorithm leads to less political homogeneity within the neighborhood than both the comparison group and networks grown through social endorsement. Structural and ideological echo chambers are not synonymous. This study highlights a need to distinguish between structural and homogeneity-based criteria across research interested in quantifying echo chambers and their impacts online.  

\subsection*{Amplification Paradox}
Researchers have noted that while recommendation systems surely impact user actions, in isolation they do not explain interaction with extreme content \cite{Ribeiro2023-kc,Chen2022-bh,Hosseinmardi2021-nc}. These studies focus on content-recommendation on YouTube but we find congruent results for the friend recommendation algorithm on Twitter. 

We find that neither the purely recommender system driven nor social-endorsement based strategies of network growth create networks that perfectly match the comparison Twitter users' personal networks. The comparison groups more closely resemble the RS networks in terms of the popularity of the alters but is more similar to the SE grown networks in terms of the number of friends the alters have. Measures of frequency of verification, activity level, and partisanship all differ between the three groups compared here. In terms of network homogeneity and exposure to false and misleading narratives, the comparison groups fall between the values of the RS and SE networks. This could suggest that while the algorithm plays a role in shaping personal networks, the interaction between algorithmic suggestion and user choice is imperative in explaining echo chambers observed online, aligning with \cite{Ribeiro2023-kc}.

We observe that the strongest partisan echo chambers result from Republican-seeded audit accounts that grew their networks through social-endorsement. These accounts had the least ideologically diverse neighborhoods and saw the most tweets about false or misleading election narratives. The Republican comparison groups had similarly high exposure to questionable election narratives. These differences across political parties are consistent with previous work demonstrating that right leaning users were more likely to engage with misinformation online \cite{kennedy2022repeat,Nikolov2021-rh}. For the Republican-seeded accounts, the friend recommendation system seems to mitigate potential exposure to those narratives and, for both left- and right-seeded accounts, produces less politically homogeneous networks. This seems to indicate that the algorithm on its own is not the key driver of the echo chambers that other research has shown exist on platforms. 

\subsection*{Implications}
The results of this work contribute to understanding \textit{what} the outcomes of the recommendation system are but cannot illuminate \textit{how} they achieve these outcomes or \textit{why} they were designed this way. The comparatively more diverse information ecosystems created by the recommender may be due to purposeful design choices or they may merely be artifacts of an algorithm optimized for other outcomes. It is reasonable to assume that Twitter, along with other social media platforms, optimizes for user engagement and revenue from advertisement exposure when designing these algorithms. This paper provides a roadmap for validating whether those outcomes are compatible with promoting healthy information environments. 
%We don't know how the algorithm works but we can surmise that it is angled at maximing user engagement on the platform. this only shows us the outcomes, not the why. we dont know how the algorithm works or what twitters goal is It could be that Twitter is actually trying to diversify sources, or this could be explained by simple friend-of-friend recommenders that happens to be lead to more diversity. this could be a simple artifact of an algorithm based on other factors such as completing triangles or popularity. 

\subsection*{Limitations}
A primary limitation of this study is the limited number of automated accounts. Previous audit studies have outlined the difficulty of conducting audit studies, including the logistics in creating accounts, challenges in data collection due to the volatility of HTML pages, and the ethical considerations of creating many accounts \cite{Bartley2021-we,Bandy2021-lg,Chen2021-tg}. Additionally, we initialized each audit account with a single friend when in reality it is likely that the first several friends have a strong influence on what is recommended. The stochastic choices at the beginning of the process potentially had an impact on the final resulting network. These early random choices may explain some of the more counter-intuitive results, such as the Republican-seeded accounts that ended up with majority left-leaning friends or the two Democrat-seeded accounts that ended up with few left-leaning friends. 

Another limitation is that the audit accounts did not post content, engage with content, or attract followers the way a real user might. As pointed out in \cite{Ribeiro2023-kc}, the interaction between algorithmic suggestion and user choice is non-negligible and limits the generalizability of studies such as ours, which isolate the algorithm from other factors. \new{We encourage future studies that assess the impact of engagement choices on recommendation systems.} 

We also consider the fact that this study offers only a snapshot of the algorithms and platform as they stood during October and November of 2022. Since that time there have been notable changes in the user-base, content recommendation algorithm, and availability of data on Twitter \cite{Rohlinger2023-fg,Anyanwu_Ray_2022,Chang_Deshmukh_Armsworth_Masuda_2023}. At the current moment the study described here is not replicable due to changes in Twitter's API, which is a hurdle that the community of social media researchers are still navigating. We also consider a relatively narrow case study of the U.S. 2022 midterm elections. \cameraready{While the findings of this work may not translate directly to other platforms or contexts, the methodologies developed and presented in this paper offer a starting point for future work across diverse contexts. Furthermore, the results presented here offer a baseline from which changes to the recommendation system can be measured.}

\section{Conclusion}

In this study, we paired the use of automated accounts with observational data to conduct an algorithmic audit of Twitter's friend recommendation algorithm in the context of US politics and political polarization. To the best of our knowledge, it is the first empirical audit, not a simulation, of friend recommenders on Twitter. We examined how the recommendation algorithm impacts the structural qualities, partisan makeup, and potential exposure to misinformation for users new to the platform during the 2022 US midterm elections. Our findings indicate that the friend recommendation algorithm leads accounts to densely connected neighborhoods that are less politically focused and more ideologically diverse than when networks are grown through social endorsement. Furthermore, accounts recommended by the algorithm are less likely to share content about false and misleading election narratives. These insights indicate that Twitter's friend recommendation algorithm alone is not the key contributor to the political echo chambers that have been shown to exist on the platform. These findings highlight the pressing need to understand the contributing social factors to online echo chambers, particularly those on social media platforms, and how those social factors interact with algorithmic decisions to create unreliable online environments.

\subsection*{\emph{Ethical statement}.}
While designing and executing this study, considerations were taken to minimize potential harm, deception, or unintended impact. First, we choose to have the automated accounts be passive observers and took care not to impersonate real users. Each account was given a nondescript name, lacked a profile photo, and had an empty bio. The accounts never tweeted, retweeted, or otherwise shared content; they did not privately message or otherwise engage with any user on the platform outside of following accounts. The accounts also followed a maximum of 12 accounts each day, growing their network over time rather than all at once. These actions are in line with Twitter's guidelines on automated accounts including rules against spamming or misleading users, or following accounts in a "bulk, aggressive, or indiscriminate" manner~\footnote{https://help.twitter.com/en/rules-and-policies/twitter-automation}. \new{Additionally, given that political affiliation is potentially sensitive information, we perform only aggregate analysis of engagement with political content and make no attempt to link users to offline behavior such as voting records, consistent with Twitter's terms of service~\footnote{https://developer.twitter.com/en/developer-terms/more-on-restricted-use-cases}}

\begin{acks}
% Identification of funding sources and other support, and thanks to
% individuals and groups that assisted in the research and the
% preparation of the work should be included in an acknowledgment
% section, which is placed just before the reference section in your
% document.
\cameraready{Funding for this work has come from the University of Washington's Center for an Informed Public, the John S. and James L. Knight Foundation (G-2019-58788), the William and Flora Hewlett Foundation (2022-00952-GRA), the Election Trust Initiative, and the National Science Foundation (grant \#2120496) as well as NSF Graduate Research Fellowships under Grant No DGE-2140004, for both Joseph S. Schafer and Kayla Duskin. Any opinions, findings, conclusions, or recommendations expressed in this material are those of the authors and do not necessarily reflect the views of the National Science Foundation or other funders.} 

\cameraready{The authors would like to acknowledge Prerna Juneja for feedback and advice regarding algorithmic audits, Prerna Sheokand for help with data analysis ideation, as well as Lia Bozarth and Alex Loddengaard for their technical support. }

\end{acks}

\bibliographystyle{ACM-Reference-Format}
\bibliography{sources}


\begin{thebibliography}{48}

%%% the Web address.

\ifx \showCODEN    \undefined \def \showCODEN     #1{\unskip}     \fi
\ifx \showDOI      \undefined \def \showDOI       #1{#1}\fi
\ifx \showISBNx    \undefined \def \showISBNx     #1{\unskip}     \fi
\ifx \showISBNxiii \undefined \def \showISBNxiii  #1{\unskip}     \fi
\ifx \showISSN     \undefined \def \showISSN      #1{\unskip}     \fi
\ifx \showLCCN     \undefined \def \showLCCN      #1{\unskip}     \fi
\ifx \shownote     \undefined \def \shownote      #1{#1}          \fi
\ifx \showarticletitle \undefined \def \showarticletitle #1{#1}   \fi
\ifx \showURL      \undefined \def \showURL       {\relax}        \fi
% invisible to TeX
\providecommand\bibfield[2]{#2}
\providecommand\bibinfo[2]{#2}
\providecommand\natexlab[1]{#1}
\providecommand\showeprint[2][]{arXiv:#2}

\bibitem[Anyanwu and Ray(2022)]%
        {Anyanwu_Ray_2022}
\bibfield{author}{\bibinfo{person}{Joy Anyanwu} {and} \bibinfo{person}{Rashawn Ray}.} \bibinfo{year}{2022}\natexlab{}.
\newblock \bibinfo{title}{Why is Elon Musk’s Twitter takeover increasing hate speech?}
\newblock
\newblock
\urldef\tempurl%
\url{https://www.brookings.edu/articles/why-is-elon-musks-twitter-takeover-increasing-hate-speech/}
\showURL{%
\tempurl}


\bibitem[Bandy and Diakopoulos(2021)]%
        {Bandy2021-lg}
\bibfield{author}{\bibinfo{person}{Jack Bandy} {and} \bibinfo{person}{Nicholas Diakopoulos}.} \bibinfo{year}{2021}\natexlab{}.
\newblock \showarticletitle{More Accounts, Fewer Links: How Algorithmic Curation Impacts Media Exposure in Twitter Timelines}.
\newblock \bibinfo{journal}{\emph{Proc. ACM Hum.-Comput. Interact.}} \bibinfo{volume}{5}, \bibinfo{number}{CSCW1} (\bibinfo{date}{April} \bibinfo{year}{2021}), \bibinfo{pages}{1--28}.
\newblock


\bibitem[Bandy and Lazovich(2022)]%
        {Bandy_Lazovich_2022}
\bibfield{author}{\bibinfo{person}{Jack Bandy} {and} \bibinfo{person}{Tomo Lazovich}.} \bibinfo{year}{2022}\natexlab{}.
\newblock \bibinfo{title}{Exposure to Marginally Abusive Content on Twitter}.  (\bibinfo{date}{Jul} \bibinfo{year}{2022}).
\newblock
\urldef\tempurl%
\url{https://doi.org/10.2139/ssrn.4175612}
\showDOI{\tempurl}


\bibitem[Bartley et~al\mbox{.}(2021)]%
        {Bartley2021-we}
\bibfield{author}{\bibinfo{person}{Nathan Bartley}, \bibinfo{person}{Andres Abeliuk}, \bibinfo{person}{Emilio Ferrara}, {and} \bibinfo{person}{Kristina Lerman}.} \bibinfo{year}{2021}\natexlab{}.
\newblock \showarticletitle{Auditing Algorithmic Bias on Twitter}. In \bibinfo{booktitle}{\emph{13th {ACM} Web Science Conference 2021}} (Virtual Event, United Kingdom) \emph{(\bibinfo{series}{WebSci '21})}. \bibinfo{publisher}{Association for Computing Machinery}, \bibinfo{address}{New York, NY, USA}, \bibinfo{pages}{65--73}.
\newblock


\bibitem[Beers et~al\mbox{.}(2023)]%
        {Beers_Schafer_Kennedy_Wack_Spiro_Starbird_2023}
\bibfield{author}{\bibinfo{person}{Andrew Beers}, \bibinfo{person}{Joseph~S. Schafer}, \bibinfo{person}{Ian Kennedy}, \bibinfo{person}{Morgan Wack}, \bibinfo{person}{Emma~S. Spiro}, {and} \bibinfo{person}{Kate Starbird}.} \bibinfo{year}{2023}\natexlab{}.
\newblock \showarticletitle{Followback clusters, satellite audiences, and bridge nodes: Coengagement networks for the 2020 US election}.
\newblock \bibinfo{journal}{\emph{Proceedings of the International AAAI Conference on Web and Social Media}}  \bibinfo{volume}{17} (\bibinfo{date}{Jun} \bibinfo{year}{2023}), \bibinfo{pages}{59–71}.
\newblock
\showISSN{2334-0770}


\bibitem[Brown et~al\mbox{.}(2022)]%
        {Brown2022-ai}
\bibfield{author}{\bibinfo{person}{Megan~A Brown}, \bibinfo{person}{James Bisbee}, \bibinfo{person}{Angela Lai}, \bibinfo{person}{Richard Bonneau}, \bibinfo{person}{Jonathan Nagler}, {and} \bibinfo{person}{Joshua~A Tucker}.} \bibinfo{year}{2022}\natexlab{}.
\newblock \bibinfo{title}{Echo Chambers, Rabbit Holes, and Algorithmic Bias: How {YouTube} Recommends Content to Real Users}.  (\bibinfo{date}{May} \bibinfo{year}{2022}).
\newblock


\bibitem[Chang et~al\mbox{.}(2023)]%
        {Chang_Deshmukh_Armsworth_Masuda_2023}
\bibfield{author}{\bibinfo{person}{Charlotte~H. Chang}, \bibinfo{person}{Nikhil~R. Deshmukh}, \bibinfo{person}{Paul~R. Armsworth}, {and} \bibinfo{person}{Yuta~J. Masuda}.} \bibinfo{year}{2023}\natexlab{}.
\newblock \showarticletitle{Environmental users abandoned Twitter after Musk takeover}.
\newblock \bibinfo{journal}{\emph{Trends in ecology and evolution}} \bibinfo{volume}{38}, \bibinfo{number}{10} (\bibinfo{date}{Oct} \bibinfo{year}{2023}), \bibinfo{pages}{893–895}.
\newblock
\showISSN{0169-5347}


\bibitem[Chen et~al\mbox{.}(2022)]%
        {Chen2022-bh}
\bibfield{author}{\bibinfo{person}{Annie~Y Chen}, \bibinfo{person}{Brendan Nyhan}, \bibinfo{person}{Jason Reifler}, \bibinfo{person}{Ronald~E Robertson}, {and} \bibinfo{person}{Christo Wilson}.} \bibinfo{year}{2022}\natexlab{}.
\newblock \showarticletitle{Subscriptions and external links help drive resentful users to alternative and extremist {YouTube} videos}.
\newblock  (\bibinfo{date}{April} \bibinfo{year}{2022}).
\newblock
\showeprint[arxiv]{2204.10921}~[cs.SI]


\bibitem[Chen et~al\mbox{.}(2021)]%
        {Chen2021-tg}
\bibfield{author}{\bibinfo{person}{Wen Chen}, \bibinfo{person}{Diogo Pacheco}, \bibinfo{person}{Kai-Cheng Yang}, {and} \bibinfo{person}{Filippo Menczer}.} \bibinfo{year}{2021}\natexlab{}.
\newblock \showarticletitle{Neutral bots probe political bias on social media}.
\newblock \bibinfo{journal}{\emph{Nat. Commun.}} \bibinfo{volume}{12}, \bibinfo{number}{1} (\bibinfo{date}{Sept.} \bibinfo{year}{2021}), \bibinfo{pages}{5580}.
\newblock


\bibitem[Cinelli et~al\mbox{.}(2021)]%
        {Cinelli2021-jr}
\bibfield{author}{\bibinfo{person}{Matteo Cinelli}, \bibinfo{person}{Gianmarco De~Francisci~Morales}, \bibinfo{person}{Alessandro Galeazzi}, \bibinfo{person}{Walter Quattrociocchi}, {and} \bibinfo{person}{Michele Starnini}.} \bibinfo{year}{2021}\natexlab{}.
\newblock \showarticletitle{The echo chamber effect on social media}.
\newblock \bibinfo{journal}{\emph{Proc. Natl. Acad. Sci. U. S. A.}} \bibinfo{volume}{118}, \bibinfo{number}{9} (\bibinfo{date}{March} \bibinfo{year}{2021}).
\newblock


\bibitem[Cinus et~al\mbox{.}(2022)]%
        {cinus2022effect}
\bibfield{author}{\bibinfo{person}{Federico Cinus}, \bibinfo{person}{Marco Minici}, \bibinfo{person}{Corrado Monti}, {and} \bibinfo{person}{Francesco Bonchi}.} \bibinfo{year}{2022}\natexlab{}.
\newblock \showarticletitle{The effect of people recommenders on echo chambers and polarization}. In \bibinfo{booktitle}{\emph{Proceedings of the International AAAI Conference on Web and Social Media}}, Vol.~\bibinfo{volume}{16}. \bibinfo{pages}{90--101}.
\newblock


\bibitem[Cota et~al\mbox{.}(2019)]%
        {Cota2019-ut}
\bibfield{author}{\bibinfo{person}{Wesley Cota}, \bibinfo{person}{Silvio~C Ferreira}, \bibinfo{person}{Romualdo Pastor-Satorras}, {and} \bibinfo{person}{Michele Starnini}.} \bibinfo{year}{2019}\natexlab{}.
\newblock \showarticletitle{Quantifying echo chamber effects in information spreading over political communication networks}.
\newblock \bibinfo{journal}{\emph{EPJ Data Science}} \bibinfo{volume}{8}, \bibinfo{number}{1} (\bibinfo{date}{Dec.} \bibinfo{year}{2019}), \bibinfo{pages}{1--13}.
\newblock


\bibitem[Del~Vicario et~al\mbox{.}(2016)]%
        {DelVicario2016}
\bibfield{author}{\bibinfo{person}{Michela Del~Vicario}, \bibinfo{person}{Alessandro Bessi}, \bibinfo{person}{Fabiana Zollo}, \bibinfo{person}{Fabio Petroni}, \bibinfo{person}{Antonio Scala}, \bibinfo{person}{Guido Caldarelli}, \bibinfo{person}{H. Eugene~Stanley}, {and} \bibinfo{person}{Walter Quattrociocchi}.} \bibinfo{year}{2016}\natexlab{}.
\newblock \showarticletitle{The spreading of misinformation online}.
\newblock \bibinfo{journal}{\emph{Proceedings of the National Academy of Sciences of the United States of America}} \bibinfo{volume}{113}, \bibinfo{number}{3} (\bibinfo{date}{Jan} \bibinfo{year}{2016}), \bibinfo{pages}{554–559}.
\newblock
\showISSN{0027-8424}


\bibitem[Diakopoulos et~al\mbox{.}(2021)]%
        {Diakopoulos2021-dq}
\bibfield{author}{\bibinfo{person}{Nicholas Diakopoulos}, \bibinfo{person}{Jack Bandy}, {and} \bibinfo{person}{Henry Dambanemuya}.} \bibinfo{year}{2021}\natexlab{}.
\newblock \bibinfo{title}{Auditing {Human-Machine} Communication Systems Using Simulated Humans}.  (\bibinfo{date}{Aug.} \bibinfo{year}{2021}).
\newblock


\bibitem[Dubois and Blank(2018)]%
        {Dubois2018-dc}
\bibfield{author}{\bibinfo{person}{Elizabeth Dubois} {and} \bibinfo{person}{Grant Blank}.} \bibinfo{year}{2018}\natexlab{}.
\newblock \showarticletitle{The echo chamber is overstated: The moderating effect of political interest and diverse media}.
\newblock \bibinfo{journal}{\emph{Inf. Commun. Soc.}} \bibinfo{volume}{21}, \bibinfo{number}{5} (\bibinfo{date}{May} \bibinfo{year}{2018}), \bibinfo{pages}{729--745}.
\newblock


\bibitem[Ferrara et~al\mbox{.}(2022)]%
        {Ferrara2022-ui}
\bibfield{author}{\bibinfo{person}{Antonio Ferrara}, \bibinfo{person}{Lisette Espin-Noboa}, \bibinfo{person}{Fariba Karimi}, {and} \bibinfo{person}{Claudia Wagner}.} \bibinfo{year}{2022}\natexlab{}.
\newblock \showarticletitle{Link recommendations: Their impact on network structure and minorities}. In \bibinfo{booktitle}{\emph{Proceedings of the 14th {ACM} Web Science Conference 2022}} (Barcelona, Spain) \emph{(\bibinfo{series}{WebSci '22})}. \bibinfo{publisher}{Association for Computing Machinery}, \bibinfo{address}{New York, NY, USA}, \bibinfo{pages}{228--238}.
\newblock


\bibitem[Flaxman et~al\mbox{.}(2016)]%
        {Flaxman_Goel_Rao_2016}
\bibfield{author}{\bibinfo{person}{Seth Flaxman}, \bibinfo{person}{Sharad Goel}, {and} \bibinfo{person}{Justin~M. Rao}.} \bibinfo{year}{2016}\natexlab{}.
\newblock \showarticletitle{Filter Bubbles, Echo Chambers, and Online News Consumption}.
\newblock \bibinfo{journal}{\emph{Public opinion quarterly}} \bibinfo{volume}{80}, \bibinfo{number}{S1} (\bibinfo{date}{Mar} \bibinfo{year}{2016}), \bibinfo{pages}{298–320}.
\newblock
\showISSN{0033-362X}


\bibitem[Garimella et~al\mbox{.}(2018)]%
        {Garimella2018}
\bibfield{author}{\bibinfo{person}{Kiran Garimella}, \bibinfo{person}{Gianmarco De~Francisci~Morales}, \bibinfo{person}{Aristides Gionis}, {and} \bibinfo{person}{Michael Mathioudakis}.} \bibinfo{year}{2018}\natexlab{}.
\newblock \showarticletitle{Political Discourse on Social Media: Echo Chambers, Gatekeepers, and the Price of Bipartisanship}. In \bibinfo{booktitle}{\emph{Proceedings of the 2018 World Wide Web Conference}} \emph{(\bibinfo{series}{WWW ’18})}. \bibinfo{publisher}{International World Wide Web Conferences Steering Committee}, \bibinfo{address}{Republic and Canton of Geneva, CHE}, \bibinfo{pages}{913–922}.
\newblock
\showISBNx{9781450356398}


\bibitem[Garimella and Weber(2017)]%
        {Garimella2017-om}
\bibfield{author}{\bibinfo{person}{Venkata Rama~Kiran Garimella} {and} \bibinfo{person}{Ingmar Weber}.} \bibinfo{year}{2017}\natexlab{}.
\newblock \showarticletitle{A {Long-Term} Analysis of Polarization on Twitter}.
\newblock \bibinfo{journal}{\emph{ICWSM}} \bibinfo{volume}{11}, \bibinfo{number}{1} (\bibinfo{date}{May} \bibinfo{year}{2017}), \bibinfo{pages}{528--531}.
\newblock


\bibitem[Garrett(2009)]%
        {Garrett_2009}
\bibfield{author}{\bibinfo{person}{R.~Kelly Garrett}.} \bibinfo{year}{2009}\natexlab{}.
\newblock \showarticletitle{Echo chambers online?: Politically motivated selective exposure among Internet news users}.
\newblock \bibinfo{journal}{\emph{Journal of computer-mediated communication: JCMC}} \bibinfo{volume}{14}, \bibinfo{number}{2} (\bibinfo{date}{Jan} \bibinfo{year}{2009}), \bibinfo{pages}{265–285}.
\newblock


\bibitem[Geurkink(2023)]%
        {wiredTwittersOpen}
\bibfield{author}{\bibinfo{person}{Brandi Geurkink}.} \bibinfo{year}{2023}\natexlab{}.
\newblock \showarticletitle{Twitter’s Open Source Algorithm Is a Red Herring}.
\newblock \bibinfo{howpublished}{https://www.wired.com/story/twitters-open-source-algorithm-is-a-red-herring/}.
\newblock \bibinfo{journal}{\emph{Wired}} (\bibinfo{date}{7 April} \bibinfo{year}{2023}).
\newblock


\bibitem[Guess et~al\mbox{.}(2018)]%
        {Guess_2018}
\bibfield{author}{\bibinfo{person}{Andrew Guess}, \bibinfo{person}{Benjamin Lyons}, \bibinfo{person}{Brendan Nyhan}, {and} \bibinfo{person}{Jason Reifler}.} \bibinfo{year}{2018}\natexlab{}.
\newblock \showarticletitle{Avoiding the echo chamber about echo chambers: Why selective exposure to like-minded political news is less prevalent than you think}.
\newblock  (\bibinfo{date}{Jan.} \bibinfo{year}{2018}).
\newblock


\bibitem[Haroon et~al\mbox{.}(2022)]%
        {Haroon2022-ks}
\bibfield{author}{\bibinfo{person}{Muhammad Haroon}, \bibinfo{person}{Anshuman Chhabra}, \bibinfo{person}{Xin Liu}, \bibinfo{person}{Prasant Mohapatra}, \bibinfo{person}{Zubair Shafiq}, {and} \bibinfo{person}{Magdalena Wojcieszak}.} \bibinfo{year}{2022}\natexlab{}.
\newblock \showarticletitle{{YouTube}, The Great Radicalizer? Auditing and Mitigating Ideological Biases in {YouTube} Recommendations}.
\newblock  (\bibinfo{date}{March} \bibinfo{year}{2022}).
\newblock
\showeprint[arxiv]{2203.10666}~[cs.CY]


\bibitem[Hosseinmardi et~al\mbox{.}(2021)]%
        {Hosseinmardi2021-nc}
\bibfield{author}{\bibinfo{person}{Homa Hosseinmardi}, \bibinfo{person}{Amir Ghasemian}, \bibinfo{person}{Aaron Clauset}, \bibinfo{person}{Markus Mobius}, \bibinfo{person}{David~M Rothschild}, {and} \bibinfo{person}{Duncan~J Watts}.} \bibinfo{year}{2021}\natexlab{}.
\newblock \showarticletitle{Examining the consumption of radical content on {YouTube}}.
\newblock \bibinfo{journal}{\emph{Proceedings of the National Academy of Sciences}} \bibinfo{volume}{118}, \bibinfo{number}{32} (\bibinfo{year}{2021}), \bibinfo{pages}{e2101967118}.
\newblock


\bibitem[Hussein et~al\mbox{.}(2020)]%
        {Hussein2020-tu}
\bibfield{author}{\bibinfo{person}{Eslam Hussein}, \bibinfo{person}{Prerna Juneja}, {and} \bibinfo{person}{Tanushree Mitra}.} \bibinfo{year}{2020}\natexlab{}.
\newblock \showarticletitle{Measuring Misinformation in Video Search Platforms: An Audit Study on {YouTube}}.
\newblock \bibinfo{journal}{\emph{Proc. ACM Hum.-Comput. Interact.}} \bibinfo{volume}{4}, \bibinfo{number}{CSCW1} (\bibinfo{date}{May} \bibinfo{year}{2020}), \bibinfo{pages}{1--27}.
\newblock


\bibitem[Kennedy et~al\mbox{.}(2022)]%
        {kennedy2022repeat}
\bibfield{author}{\bibinfo{person}{Ian Kennedy}, \bibinfo{person}{Morgan Wack}, \bibinfo{person}{Andrew Beers}, \bibinfo{person}{Joseph~S Schafer}, \bibinfo{person}{Isabella Garcia-Camargo}, \bibinfo{person}{Emma~S Spiro}, {and} \bibinfo{person}{Kate Starbird}.} \bibinfo{year}{2022}\natexlab{}.
\newblock \showarticletitle{Repeat Spreaders and Election Delegitimization: A Comprehensive Dataset of Misinformation Tweets from the 2020 US Election}.
\newblock \bibinfo{journal}{\emph{Journal of Quantitative Description: Digital Media}}  \bibinfo{volume}{2} (\bibinfo{year}{2022}).
\newblock


\bibitem[Krackhardt and Stern(1988)]%
        {krackhardt1988informal}
\bibfield{author}{\bibinfo{person}{David Krackhardt} {and} \bibinfo{person}{Robert~N Stern}.} \bibinfo{year}{1988}\natexlab{}.
\newblock \showarticletitle{Informal networks and organizational crises: An experimental simulation}.
\newblock \bibinfo{journal}{\emph{Social psychology quarterly}} (\bibinfo{year}{1988}), \bibinfo{pages}{123--140}.
\newblock


\bibitem[Kubin and von Sikorski(2021)]%
        {Kubin2021-lc}
\bibfield{author}{\bibinfo{person}{Emily Kubin} {and} \bibinfo{person}{Christian von Sikorski}.} \bibinfo{year}{2021}\natexlab{}.
\newblock \showarticletitle{The role of (social) media in political polarization: a systematic review}.
\newblock \bibinfo{journal}{\emph{Annals of the International Communication Association}} \bibinfo{volume}{45}, \bibinfo{number}{3} (\bibinfo{date}{July} \bibinfo{year}{2021}), \bibinfo{pages}{188--206}.
\newblock


\bibitem[Lee et~al\mbox{.}(2014)]%
        {Lee2014-hh}
\bibfield{author}{\bibinfo{person}{Jae~Kook Lee}, \bibinfo{person}{Jihyang Choi}, \bibinfo{person}{Cheonsoo Kim}, {and} \bibinfo{person}{Yonghwan Kim}.} \bibinfo{year}{2014}\natexlab{}.
\newblock \showarticletitle{Social Media, Network Heterogeneity, and Opinion Polarization}.
\newblock \bibinfo{journal}{\emph{J. Commun.}} \bibinfo{volume}{64}, \bibinfo{number}{4} (\bibinfo{date}{Aug.} \bibinfo{year}{2014}), \bibinfo{pages}{702--722}.
\newblock


\bibitem[Lee et~al\mbox{.}(2022)]%
        {Lee_Sun_Jang_Connelly_2022}
\bibfield{author}{\bibinfo{person}{Sun~Kyong Lee}, \bibinfo{person}{Juhyung Sun}, \bibinfo{person}{Seulki Jang}, {and} \bibinfo{person}{Shane Connelly}.} \bibinfo{year}{2022}\natexlab{}.
\newblock \showarticletitle{Misinformation of COVID-19 vaccines and vaccine hesitancy}.
\newblock \bibinfo{journal}{\emph{Scientific reports}} \bibinfo{volume}{12}, \bibinfo{number}{1} (\bibinfo{date}{Aug} \bibinfo{year}{2022}), \bibinfo{pages}{13681}.
\newblock
\showISSN{2045-2322}


\bibitem[McClain et~al\mbox{.}(2021)]%
        {mcclain2021behaviors}
\bibfield{author}{\bibinfo{person}{Colleen McClain}, \bibinfo{person}{Regina Widjaya}, \bibinfo{person}{Gonzalo Rivero}, {and} \bibinfo{person}{Aaron Smith}.} \bibinfo{year}{2021}\natexlab{}.
\newblock \showarticletitle{The behaviors and attitudes of US adults on Twitter}.
\newblock  (\bibinfo{year}{2021}).
\newblock


\bibitem[Nikolov et~al\mbox{.}(2021a)]%
        {Nikolov_Flammini_Menczer_2021}
\bibfield{author}{\bibinfo{person}{Dimitar Nikolov}, \bibinfo{person}{Alessandro Flammini}, {and} \bibinfo{person}{Filippo Menczer}.} \bibinfo{year}{2021}\natexlab{a}.
\newblock \showarticletitle{Right and left, partisanship predicts (asymmetric) vulnerability to misinformation}.
\newblock \bibinfo{journal}{\emph{Harvard Kennedy School Misinformation Review}} (\bibinfo{date}{Feb} \bibinfo{year}{2021}).
\newblock
\urldef\tempurl%
\url{https://doi.org/10.37016/mr-2020-55}
\showDOI{\tempurl}


\bibitem[Nikolov et~al\mbox{.}(2021b)]%
        {Nikolov2021-rh}
\bibfield{author}{\bibinfo{person}{Dimitar Nikolov}, \bibinfo{person}{Alessandro Flammini}, {and} \bibinfo{person}{Filippo Menczer}.} \bibinfo{year}{2021}\natexlab{b}.
\newblock \showarticletitle{Right and left, partisanship predicts (asymmetric) vulnerability to misinformation}.
\newblock \bibinfo{journal}{\emph{HKS Misinfo Review}} (\bibinfo{date}{Feb.} \bibinfo{year}{2021}).
\newblock


\bibitem[Nyce(2023)]%
        {atlanticTransperent}
\bibfield{author}{\bibinfo{person}{Caroline~Mimbs Nyce}.} \bibinfo{year}{2023}\natexlab{}.
\newblock \showarticletitle{The New ‘Transparent’ Twitter Isn’t Very Transparent}.
\newblock \bibinfo{howpublished}{https://www.theatlantic.com/technology/archive/2023/04/elon-musk-twitter-algorithm-code-public/673642/}.
\newblock \bibinfo{journal}{\emph{The Atlantic}} (\bibinfo{date}{5 April} \bibinfo{year}{2023}).
\newblock


\bibitem[Papadamou et~al\mbox{.}(2022)]%
        {Papadamou2022-ym}
\bibfield{author}{\bibinfo{person}{Kostantinos Papadamou}, \bibinfo{person}{Savvas Zannettou}, \bibinfo{person}{Jeremy Blackburn}, \bibinfo{person}{Emiliano~De Cristofaro}, \bibinfo{person}{Gianluca Stringhini}, {and} \bibinfo{person}{Michael Sirivianos}.} \bibinfo{year}{2022}\natexlab{}.
\newblock \showarticletitle{``it is just a flu'': Assessing the effect of watch history on {YouTube's} pseudoscientific video recommendations}.
\newblock \bibinfo{journal}{\emph{Proceedings of the International AAAI Conference on Web and Social Media}}  \bibinfo{volume}{16} (\bibinfo{date}{May} \bibinfo{year}{2022}), \bibinfo{pages}{723--734}.
\newblock


\bibitem[Pariser(2011)]%
        {pariser2011filter}
\bibfield{author}{\bibinfo{person}{Eli Pariser}.} \bibinfo{year}{2011}\natexlab{}.
\newblock \bibinfo{booktitle}{\emph{The filter bubble: What the Internet is hiding from you}}.
\newblock \bibinfo{publisher}{penguin UK}.
\newblock


\bibitem[{Pew Research Center}(2017)]%
        {pew_misinfo}
\bibfield{author}{\bibinfo{person}{{Pew Research Center}}.} \bibinfo{year}{2017}\natexlab{}.
\newblock \bibinfo{title}{The Future of Truth and Misinformation Online}.
\newblock
\newblock
\urldef\tempurl%
\url{https://www.pewresearch.org/internet/2017/10/19/the-future-of-truth-and-misinformation-online/}
\showURL{%
\tempurl}


\bibitem[{Pew Research Center}(2022)]%
        {pew_2022}
\bibfield{author}{\bibinfo{person}{{Pew Research Center}}.} \bibinfo{year}{2022}\natexlab{}.
\newblock \bibinfo{title}{Social Media and News Fact Sheet}.
\newblock
\newblock
\urldef\tempurl%
\url{https://www.pewresearch.org/journalism/fact-sheet/social-media-and-news-fact-sheet/}
\showURL{%
\tempurl}


\bibitem[Ribeiro et~al\mbox{.}(2023)]%
        {Ribeiro2023-kc}
\bibfield{author}{\bibinfo{person}{Manoel~Horta Ribeiro}, \bibinfo{person}{Veniamin Veselovsky}, {and} \bibinfo{person}{Robert West}.} \bibinfo{year}{2023}\natexlab{}.
\newblock \showarticletitle{The Amplification Paradox in Recommender Systems}.
\newblock  (\bibinfo{date}{Feb.} \bibinfo{year}{2023}).
\newblock
\showeprint[arxiv]{2302.11225}~[cs.CY]


\bibitem[Rohlinger et~al\mbox{.}(2023)]%
        {Rohlinger2023-fg}
\bibfield{author}{\bibinfo{person}{Deana~A Rohlinger}, \bibinfo{person}{Kyle Rose}, \bibinfo{person}{Sarah Warren}, {and} \bibinfo{person}{Stuart Shulman}.} \bibinfo{year}{2023}\natexlab{}.
\newblock \showarticletitle{Does the Musk Twitter Takeover Matter? Political Influencers, Their Arguments, and the Quality of Information They Share}.
\newblock \bibinfo{journal}{\emph{Socius}}  \bibinfo{volume}{9} (\bibinfo{date}{Jan.} \bibinfo{year}{2023}), \bibinfo{pages}{23780231231152193}.
\newblock


\bibitem[Sanchez and Middlemass(2022)]%
        {Sanchez_Middlemass_2022}
\bibfield{author}{\bibinfo{person}{Gabriel~R. Sanchez} {and} \bibinfo{person}{Keesha Middlemass}.} \bibinfo{year}{2022}\natexlab{}.
\newblock \showarticletitle{Misinformation is eroding the public’s confidence in democracy}.
\newblock \bibinfo{journal}{\emph{Brookings Institute}} (\bibinfo{date}{Jul} \bibinfo{year}{2022}).
\newblock
\urldef\tempurl%
\url{https://www.brookings.edu/articles/misinformation-is-eroding-the-publics-confidence-in-democracy/}
\showURL{%
\tempurl}


\bibitem[Sandvig et~al\mbox{.}(2014)]%
        {sandvig2014auditing}
\bibfield{author}{\bibinfo{person}{Christian Sandvig}, \bibinfo{person}{Kevin Hamilton}, \bibinfo{person}{Karrie Karahalios}, {and} \bibinfo{person}{Cedric Langbort}.} \bibinfo{year}{2014}\natexlab{}.
\newblock \showarticletitle{Auditing algorithms: Research methods for detecting discrimination on internet platforms}.
\newblock \bibinfo{journal}{\emph{Data and discrimination: converting critical concerns into productive inquiry}} \bibinfo{volume}{22}, \bibinfo{number}{2014} (\bibinfo{year}{2014}), \bibinfo{pages}{4349--4357}.
\newblock


\bibitem[Santos et~al\mbox{.}(2021)]%
        {Santos2021-vm}
\bibfield{author}{\bibinfo{person}{Fernando~P Santos}, \bibinfo{person}{Yphtach Lelkes}, {and} \bibinfo{person}{Simon~A Levin}.} \bibinfo{year}{2021}\natexlab{}.
\newblock \showarticletitle{Link recommendation algorithms and dynamics of polarization in online social networks}.
\newblock \bibinfo{journal}{\emph{Proc. Natl. Acad. Sci. U. S. A.}} \bibinfo{volume}{118}, \bibinfo{number}{50} (\bibinfo{date}{Dec.} \bibinfo{year}{2021}).
\newblock


\bibitem[Schafer et~al\mbox{.}(2024)]%
        {Schafer2024-dl}
\bibfield{author}{\bibinfo{person}{Joseph~S Schafer}, \bibinfo{person}{Kayla Duskin}, \bibinfo{person}{Morgan Wack}, \bibinfo{person}{Anna Beers}, \bibinfo{person}{Lia Bozarth}, \bibinfo{person}{Taylor Agajanian}, \bibinfo{person}{Mike Caulfield}, \bibinfo{person}{Emma~S Spiro}, {and} \bibinfo{person}{Kate Starbird}.} \bibinfo{year}{2024}\natexlab{}.
\newblock \bibinfo{title}{{ElectionRumors2022}: A dataset of election rumors on Twitter during the 2022 {US} midterms}.
\newblock
\newblock


\bibitem[Su et~al\mbox{.}(2016)]%
        {Su2016-ry}
\bibfield{author}{\bibinfo{person}{Jessica Su}, \bibinfo{person}{Aneesh Sharma}, {and} \bibinfo{person}{Sharad Goel}.} \bibinfo{year}{2016}\natexlab{}.
\newblock \showarticletitle{The Effect of Recommendations on Network Structure}. In \bibinfo{booktitle}{\emph{Proceedings of the 25th International Conference on World Wide Web}} (Montr{\'e}al, Qu{\'e}bec, Canada) \emph{(\bibinfo{series}{WWW '16})}. \bibinfo{publisher}{International World Wide Web Conferences Steering Committee}, \bibinfo{address}{Republic and Canton of Geneva, CHE}, \bibinfo{pages}{1157--1167}.
\newblock


\bibitem[Sunstein(2002)]%
        {Sunstein_2002}
\bibfield{author}{\bibinfo{person}{Cass~R. Sunstein}.} \bibinfo{year}{2002}\natexlab{}.
\newblock \showarticletitle{The law of group polarization}.
\newblock \bibinfo{journal}{\emph{The journal of political philosophy}} \bibinfo{volume}{10}, \bibinfo{number}{2} (\bibinfo{date}{Jun} \bibinfo{year}{2002}), \bibinfo{pages}{175–195}.
\newblock
\showISSN{0963-8016}


\bibitem[Terren and Borge-Bravo(2021)]%
        {Terren2021-fj}
\bibfield{author}{\bibinfo{person}{Ludovic Terren} {and} \bibinfo{person}{Rosa Borge-Bravo}.} \bibinfo{year}{2021}\natexlab{}.
\newblock \showarticletitle{Echo Chambers on Social Media: A Systematic Review of the Literature}.
\newblock \bibinfo{journal}{\emph{RCR}}  \bibinfo{volume}{9} (\bibinfo{date}{March} \bibinfo{year}{2021}), \bibinfo{pages}{99--118}.
\newblock


\bibitem[Tommasel and Menczer(2022)]%
        {Tommasel2022-uv}
\bibfield{author}{\bibinfo{person}{Antonela Tommasel} {and} \bibinfo{person}{Filippo Menczer}.} \bibinfo{year}{2022}\natexlab{}.
\newblock \showarticletitle{Do Recommender Systems Make Social Media More Susceptible to Misinformation Spreaders?}. In \bibinfo{booktitle}{\emph{Proceedings of the 16th {ACM} Conference on Recommender Systems}} (Seattle, WA, USA) \emph{(\bibinfo{series}{RecSys '22})}. \bibinfo{publisher}{Association for Computing Machinery}, \bibinfo{address}{New York, NY, USA}, \bibinfo{pages}{550--555}.
\newblock


\end{thebibliography}

\appendix

\end{document}